\documentclass[floats,floatfix,showpacs,amssymb,prd,twocolumn,groupedaddress,nofootinbib]{revtex4-1}

\usepackage{amssymb,amsmath,verbatim,mathtools,needspace,enumitem,etoolbox,graphicx,physics,microtype,afterpage,bm}

\usepackage[dvipsnames, usenames]{xcolor}
\usepackage[normalem]{ulem}
\usepackage{soul}

\definecolor{linkcolor}{rgb}{0.0,0.3,0.5}
\usepackage[unicode, colorlinks=true, linkcolor=linkcolor, citecolor=linkcolor, filecolor=linkcolor,urlcolor=linkcolor, pdfusetitle]{hyperref}
\usepackage[all]{hypcap}
\usepackage[T1]{fontenc}
\usepackage[utf8]{inputenc}
\usepackage{tabularx}
\usepackage{float}
\usepackage{placeins}
\usepackage{booktabs}

\allowdisplaybreaks
\usepackage{tensor}

\def\a{\alpha}
\def\b{\beta}

\newcommand{\GeV}{\; \mathrm{GeV}}

\newcommand{\be}{\begin{equation}} 
\newcommand{\ee}{\end{equation}}
\newcommand{\beq}{\begin{equation}} 
\newcommand{\eeq}{\end{equation}}
\newcommand{\bea}{\begin{equation}\begin{aligned}} 
\newcommand{\eea}{\end{aligned}\end{equation}}
\newcommand{\ba}{\begin{eqnarray}}
\newcommand{\ea}{\end{eqnarray}}

\newcommand{\bfrac}[2]{\left(\frac{#1}{#2}\right)}
\newcommand{\brac}[1]{\left( #1 \right)}
\newcommand{\mpl}{M_{\rm Pl}}

\usepackage{orcidlink}

\usepackage{csquotes}
\definecolor{tclr}{RGB}{103,103,246}

\begin{document}

\title{Generalized uncertainty principle and neutrino phenomenology}

\author{Ioannis~D.~Gialamas\orcidlink{0000-0002-2957-5276}}%~\orcid{0000-0002-2957-5276}}
\email{ioannis.gialamas@kbfi.ee}
\affiliation{Laboratory of High Energy and Computational Physics, 
National Institute of Chemical Physics and Biophysics, R{\"a}vala pst.~10, Tallinn, 10143, Estonia}
\author{Timo~J.~K{\"a}rkk{\"a}inen\orcidlink{0000-0002-2885-2235}}%~\orcid{0000-0002-2957-5276}}
\email{karkkainen@kbfi.ee}
\affiliation{Laboratory of High Energy and Computational Physics, 
National Institute of Chemical Physics and Biophysics, R{\"a}vala pst.~10, Tallinn, 10143, Estonia}
\author{Luca Marzola\orcidlink{0000-0003-2045-1100}}%~\orcid{0000-0002-2957-5276}}
\email{luca.marzola@cern.ch}

\affiliation{Laboratory of High Energy and Computational Physics, 
National Institute of Chemical Physics and Biophysics, R{\"a}vala pst.~10, Tallinn, 10143, Estonia\\}
\affiliation{Institute of Computer Science, University of Tartu, 
Narva mnt 18, 51009 Tartu, Estonia}

\begin{abstract}
\noindent
\subsection*{Abstract}
Generalized uncertainty principles are effective changes to the Heisenberg uncertainty principle that emerge in several quantum gravity models. In the present letter, we study the consequences that two classes of these modifications yield on the physics of neutrinos. Besides analyzing the change in the oscillation probabilities that the generalized uncertainty principles entail, we assess their impact on the neutrino coherence length and their possible interpretation as nonstandard neutrino interactions. Constraints cast by present and planned neutrino experiments on the generalized uncertainty principles parameters are also derived.       
\vspace{1cm}

\end{abstract}

\maketitle

\section{Introduction}

Research into the phenomenology of Generalized Uncertainty Principles (GUPs) originated with the attempt to construct a coherent theory of quantum gravity (QG)---still one of the most important goals of contemporary theoretical physics. Despite extensive research in the last decades, any of the proposed QG theories has yet to receive a first experimental validation as, generally, current data does not yet probe the regime where the proposed flavour of quantumness appears. The usual treatment of gravitational effects then commonly relies on effective classical theories, believed to hold at energy scales below the Planck mass, $\mpl\simeq 10^{19} \GeV$, where the quantum regime should finally take over. Various QG theories, including promising candidates such as string theory~\cite{Amati:1987wq,Konishi:1989wk} and loop quantum gravity~\cite{Rovelli:1994ge}, propose that the onset of a quantum regime is accompanied by the emergence of a minimum length comparable to the Planck scale $\sim 10^{-35}\, {\rm m}$  \footnote{Importantly, theories incorporating extra dimensions may support larger gravitational constant values, thereby leading to a larger Planck length and, consequently, a greater expected minimum length scale~\cite{Arkani-Hamed:1998jmv,Antoniadis:1998ig,Arkani-Hamed:1998sfv}}. Regardless of the specifics of the framework, the introduction of a minimum length requires adjustments to the Heisenberg Uncertainty Principle which, in turn, yield a GUP~\cite{Maggiore:1993rv,Garay:1994en,Kempf:1994su,Adler:1999bu,Chen:2002tu,Das:2008kaa,Adler:2010wf}. 

Since their first appearance in QG theories, GUPs have been the subject of dedicated studies aimed at exploring the related phenomenological consequences within gravity theories~\cite{Maggiore:1993zu, Maggiore:1993kv, Scardigli:1999jh}, as well as in many other fields of physics including, for instance, black holes~\cite{Adler:2001vs,Myung:2006qr,Nozari:2008gp,Bargueno:2015tea,Carr:2015nqa,Casadio:2017sze,Alonso-Serrano:2018ycq,Ong:2018zqn,Gangopadhyay:2018hhw,Mureika:2018gxl,Knipfer:2019pgi,Carr:2020hiz,Zhou:2020gww,Carr:2022ndy} and gravitational waves physics~\cite{Feng:2016tyt,Bhattacharyya:2020ooz,Diab:2020jcl,Das:2021lrb,Das:2022hjp}, with still continuing activity~\cite{Benczik:2005bh,Bambi:2007ty,Marin:2013pga,Scardigli:2014qka,Scardigli:2016pjs,Feng:2020ams,Feng:2022gdz,Banerjee:2022slh,Bernaldez:2022muh,Pachol:2023bkv,Artigas:2023lxm,Kozak:2023vlj,Bosso:2023fnb,Wojnar:2023bvv,Ali:2024tbd,Segreto:2024vtu,Hong:2024xog}. We refer the reader to Ref.~\cite{Bosso:2023aht} for a recent review on the topic. 

In this letter we continue the exploration of GUPs by investigating a possible low energy phenomenology that yields testable effects on the properties of neutrinos. These particles provide an ideal laboratory to test GUP scenarios because the flavor oscillations observed at dedicated  experiments are inherently a quantum phenomenon and, as such, are directly sensitive to the workings of quantum mechanics that GUPs modify.\footnote{A complementary analysis which uses neutrinos to test General Relativity and its extensions at the classical and semi-classical level is presented in Ref.~\cite{Luciano:2020xox}.} The work improves on previous studies of GUPs in neutrino phenomenology~\cite{Sprenger:2010dg,Torri:2024jwc} by considering different GUP formulations and updated results of neutrino oscillation experiments. The assessment of the impact of GUPs on the  neutrino decoherence length and their interpretation as nonstandard neutrino interactions are also  novel results of the analysis. In the following, prior to showing how GUPs modify the phenomenology of neutrinos, we briefly review the GUP scenarios that will be considered in our study.

\section{The generalized uncertainty principle}
% \noindent

Depending on the specific parametrization, modifications of the Heisenberg uncertainty principle introduce in the theory a minimum measurable length which can be accompanied by a maximum measurable momentum. In the following, we briefly review each of these prototypical cases which we later use in our analysis.

\subsection*{GUP-I: minimum measurable length}

The first GUP parametrization we consider is of the form~\cite{Kempf:1994su}:
\begin{align}
\label{eq:GUP}
\Delta x_i \Delta p_i \geq \frac{\hbar}{2} \big[&1+\beta \left( (\Delta p)^2 + \langle p\rangle^2 \right) \nonumber
\\ & +2\beta \left(  (\Delta p_i)^2 + \langle p_i\rangle^2 \right) \big]\,,
\end{align}
with $i=1,2,3$ and $p = \sqrt{ \sum_i p_i p_i}$. The parameter $\beta$ is dubbed \textit{GUP parameter}, and it is defined as
$\b = \b_0 l_{\rm Pl}^2/\hbar^2 = \b_0/(c M_{\rm Pl})^2$.
In this context, $\b_0$ is a dimensionless parameter that regulates the value of the minimum length while $l_{\rm Pl} $ and $M_{\rm Pl}$ are the Planck length and mass, respectively. In natural units, the GUP parameter is $\b  \simeq 7 \b_0 \times10^{-39} \GeV^{-2}$.

Upon minimizing~\eqref{eq:GUP} with respect to $\Delta p_i$, it becomes evident that the GUP implies a minimum length $\Delta x_i ^{\rm min} \sim \sqrt{\b_0} l_{\rm Pl}$ for the measurement of any spatial Cartesian coordinate. 

The GUP form under consideration arises from modifications of the Heisenberg algebra, as elaborated in~\cite{Kempf:1994su}. This altered algebra, which serves as the foundation for the GUP, is expressed by the equation:
\be
\label{eq:Heis_alg}
[x_i, p_j] = i\hbar \left( \delta_{ij} +\beta \delta_{ij} p^2 +2\beta p_i p_j \right)\,.
\ee
It is worth noting that this particular form of the commutator ensures that $[x_i, x_j] = [p_i, p_j] = 0$. We remark that this property also applies to the GUP formulation which we will discuss subsequently. 

We use the transformations
\be
\label{eq:tilded}
x_i = \tilde{x}_i\,, \qquad p_i = \tilde{p}_i (1+\beta \tilde{p}^2 ) \,, \qquad E/c = p_0 =\tilde{p}_0\,,
\ee
to define tilted variables that satisfy the canonical commutation relations  $[\tilde{x}_i, \tilde{p}_j]= i \hbar \delta_{ij}$. The squared $4-$momentum of the physical variables $p$ and $p_0$ is expanded as\footnote{Note that we have used the mostly minus sign convention and from now on we set $c=1$. }
\be
\label{eq:4mom}
    p_0^2 -p^2  = \tilde{p}_0^2 - \tilde{p}^2\left( 1 + 2\b\tilde{p}^2 +\mathcal{O}(\b^2) \right)\,,
\ee
where the higher order terms in the $\b-$expansion are Planck suppressed.  The tilded variables satisfy the canonical dispersion relation, namely $\tilde{p}_0^2 - \tilde{p}^2 = m^2$, so using this along with the inverse transformation $\tilde{p}^2 = p^2\left(1-2\b p^2 + \mathcal{O}(\b^2)\right)$, Eq.~\eqref{eq:4mom} can be rewritten as
\be
\label{eq:disp_GUP}
E^2 =m^2 +p^2(1-2\b p^2)\,,
\ee
where for $\b \rightarrow 0$  the standard energy-momentum dispersion relation is restored. At the leading order in the GUP parameter we have
\be
\label{eq:DR1}
\Delta p = \Delta E +3 \beta E^2 \Delta E +\mathcal{O}(\b^2)\,.
\ee

\subsection*{GUP-II: minimum measurable length and maximum measurable momentum}

Next, we examine the generalization suggested in~\cite{Ali:2009zq,Das:2010zf}, wherein the commutation relation satisfied by the modified position and momentum is given by
\be
 \label{eq:Heis_alg_2}
[x_i, p_j] = i\hbar \left( \delta_{ij} -\a\left(p \delta_{ij} +\frac{p_i p_j}{p} +\a^2 \left( p^2 \delta_{ij} +3p_i p_j \right) \right) \right)\,,   
\ee
where $\a = \a_0 l_{\rm Pl}/\hbar$ is the ``GUP-II parameter''. This formulation of GUP implies a minimum measurable length $\Delta x_i ^{\rm min} \sim \a_0 l_{\rm Pl}$  and a maximum measurable momentum $\Delta p_i ^{\rm max} \sim M_{\rm Pl}/\a_0$. The corresponding dispersion relation is\footnote{The momentum transformation \eqref{eq:tilded} in this case takes the form $p_i = \tilde{p}_i (1-\alpha \tilde{p}_i +2\alpha^2 \tilde{p}_i^2)$. This yields a modified dispersion relation of the form $E^2 = m^2 + p^2(1\pm \alpha p)^2$. In our analysis, we retain only the minus sign, as the alternative leads to superluminal speed of gravitational waves \cite{Das:2021lrb}.  }
\be
\label{eq:disp_GUP_2}
E^2 =m^2 +p^2(1-\a p)^2\,.
\ee
Expanding again at the leading order in the relevant GUP parameter we find
\be
\label{eq:DR2}
\Delta p = \Delta E +2 \a E \Delta E +\mathcal{O}(\a^2)\,.
\ee

Although QC models predict specific value of $\a$ or $\b$, in the following we consider arbitrary values for the parameters in order to assess the power of neutrino experiments to constrain GUP scenarios.

\section{Experimental constraints and prospects from neutrino oscillations}

Prior to quantifying the impact of the modified dispersion relations in Eqs.~\eqref{eq:disp_GUP} and~\eqref{eq:disp_GUP_2} on neutrino oscillations, we briefly summarize the workings of this effect.

\subsection{Standard neutrino oscillations}

The misalignment between neutrino mass and flavour (interaction) eigenstates is described by the Pontecorvo-Maki-Nakagawa-Sakata (PMNS) neutrino mixing matrix $U$:
\be 
\begin{pmatrix}
    \nu_e \\ \nu_\mu \\ \nu_\tau 
\end{pmatrix}
=
\begin{pmatrix}
    U_{e1} & U_{e2} & U_{e3}\\
    U_{\mu 1} & U_{\mu 2} & U_{\mu 3}\\
    U_{\tau 1} & U_{\tau 2} & U_{\tau   3}
\end{pmatrix}
\begin{pmatrix}
    \nu_1 \\ \nu_2 \\ \nu_3 
\end{pmatrix}\,.
\ee 
The matrix elements depend on three mixing angles $\theta_{12}$, $\theta_{13}$, $\theta_{23}$ and a CP violating phase $\delta$, which later we set to the experimental values provided by the \texttt{NuFit} collaboration~\cite{Esteban:2020cvm} in our numerical analysis.

Let a neutrino be produced in a pure state at $t=0$ by a Standard Model (SM) gauge interaction. The time evolution of its flavour state is given by
\be 
\nu_\ell(t) = \sum_{\substack{\ell'=e,\mu,\tau\\ i=1,2,3}} U_{\ell i}U_{\ell' i}^*e^{-iE_it/\hbar}\nu_{\ell'}(0) \equiv  \sum_{\ell'=e,\mu,\tau}A_{\ell\ell'}\nu_{\ell'}(0)\,,
\ee 
where $A_{\ell\ell'}$ is the $\ell \rightarrow \ell'$ flavour transition probability amplitude at time $t > 0$. The corresponding transition probability is
\be 
P_{\ell\ell'} =|A_{\ell\ell'}|^2 =  \sum \limits_{i,j=1}^3 U_{\ell i}^*U_{\ell' i}U_{\ell j}U_{\ell' j}^*e^{-i(E_i-E_j)t/\hbar}\,.
\ee 
As customary we assume neutrinos are ultrarelativistic, so the baseline length for neutrino oscillation is $L \approx t$. In the limit $m_i \ll p_i$, $i=1,2,3$, and utilizing the same-momentum approximation, $p_1 \approx p_2 \approx p_3 \approx p \approx E$, the oscillation phase is
\be 
\Delta_{ij} \equiv \frac{E_i-E_j}{2} = \frac{L\Delta m_{ij}^2}{4E}\,,
\ee 
where $\Delta m_{ij}^2 \equiv m_i^2-m_j^2$ are the neutrino mass squared differences. It is straightforward to show that in the two neutrino approximation with mixing angle $\theta$ and oscillation frequency $\Delta = L\Delta m_{ij}^2/4E$ the oscillation probability reduces to
\be 
P_{\ell\ell'} = \sin^2(2\theta)\sin^2\Delta,\quad \ell \neq \ell'\,.
\ee 

\subsection*{Neutrino coherence length}
    
The standard derivation of the neutrino oscillation transition probability follows from solving a purely quantum mechanical Schr\"odinger equation with a plane wave ansatz. At the leading order this produces the correct expression, but sub-leading effects should also be taken into account. Contrary to the plane wave approximation, neutrino production and detection occur in a finite, localized, spacetime region and consequently are more properly modelled in terms of wave packets which posses uncertain energy and momentum. As the neutrino mass eigenstates are non-degenerate, the wave packets will separate over time and the oscillation of the transition probability will be exponentially suppressed by a characteristic \textit{coherence length} $L_\text{coh}$ at which the wave packet separation starts to be significant. After a baseline of a few coherence lengths, the transition probabilities can be regarded as constant, signalling that decoherence took place. The expression for the transition probability computed in the wave packet formalism accounting for the effect is given by~\cite{Giunti:2007ry}
\be 
    P_{\ell\ell'} = \sum \limits_{i,j=1}^3 U_{\ell i}^*U_{\ell' i}U_{\ell j}U_{\ell' j}^*\exp\brac{-2i\Delta_{ij}-\bfrac{L}{L_{ij}^\text{coh}}^2}\,,
\ee
where the coherence length is
\be 
\label{eq:cohele}
L^\text{coh}_{ij} = \frac{4\sqrt{2}E_\nu^{2}\Delta x}{|\Delta m_{ij}^2|}\,,
\ee 
and $\Delta x$ is the neutrino position uncertainty.

\subsection*{Nonstandard neutrino interactions }

In a medium with electron number density $N_e$, the Hamiltonian regulating neutrino oscillations is given by
\be 
H = \frac{1}{2E}U^\dagger \begin{pmatrix}
    0 & 0 & 0 \\ 0 & \Delta m_{21}^2 & 0 \\ 0 & 0 & \Delta m_{31}^2
\end{pmatrix}U + \sqrt{2}G_FN_e\begin{pmatrix}
    1 & 0 & 0 \\ 0 & 0 & 0 \\ 0 & 0 & 0
\end{pmatrix}\,,
\ee 
where $G_F$ is the Fermi constant. The neutrino mixing matrices appearing in the above equation rotate the matrix hosting the squared mass differences to the flavour basis where interactions are diagonal. Theories beyond the SM, including Type II seesaw~\cite{Minkowski:1977sc,Yanagida:1979as,Gell-Mann:1979vob,Mohapatra:1979ia,Schechter:1980gr}, the Zee-Babu model~\cite{Zee:1980ai,Babu:1988ki} and $U(1)$ extensions coupling to neutrino sector~\cite{Peli:2022ybi,Trocsanyi:2018bkm,Iwamoto:2021fup,Iwamoto:2021wko,Karkkainen:2021tbh,Karkkainen:2023ozw}, can modify neutrino oscillations upon the introduction of \textit{nonstandard neutrino interactions} (NSI). These are described with dimension-6 effective operators in the Lagrangian:
\be 
\mathcal{L}_\text{NSI} = -2\sqrt{2}G_F\varepsilon^{ff',C}_{\ell\ell'}(\overline{\nu_{L,\alpha}}\gamma^\mu\nu_{L,\beta})(\overline{f}\gamma^\mu P_{C}f')\,,
\ee 
where a sum over chiralities ($C=L,R$), fermions $(f,f')$ and lepton flavours $(\ell,\ell'=e,\mu,\tau)$ is implied. The NSI parameters $\varepsilon^{ff',C}_{\ell\ell'}$ are dimensionless and in general complex numbers. Possible non-vanishing imaginary parts induce CP violation (CPV) in neutrino sector even if the standard CP phase $\delta$ vanishes. The absolute values of the NSI parameters can be interpreted as the strength of the interaction with respect to weak interaction.

Integrating out the background fermion fields, the contribution of NSI is
generally modeled in a Hamiltonian
\begin{align}
\label{eq:HNSI}
H_\text{NSI} &= \sum\limits_{f}\sqrt{2}G_FN_f
\begin{pmatrix}
    \varepsilon^f_{ee} & \varepsilon^f_{e\mu} & \varepsilon^f_{e\tau}\\
    \varepsilon^f_{\mu e} & \varepsilon^f_{\mu\mu} & \varepsilon^f_{\mu\tau}\\
    \varepsilon^f_{\tau e} & \varepsilon^f_{\tau \mu} & \varepsilon^f_{\tau\tau}\\
\end{pmatrix}\,,
\end{align}
where $N_f$ is the number density of the fermion $f$ and $\varepsilon_{\ell\ell'}^f = \varepsilon^{f,L}_{\ell\ell'}+\varepsilon^{f,R}_{\ell\ell'}$. Note that hermiticity requires $\varepsilon_{\ell\ell'}^{f,*} = \varepsilon_{\ell'\ell}^f$. The inclusion of this operator induces subleading effects in neutrino flavour transitions that modify the standard oscillation framework by changing the oscillation probability $P_{\ell\ell'}$ determined by the composition of the medium. In the following we will derive relations between NSI and GUP parameters and use the current experimental bounds on the former to constrain the latter.

\subsection{GUP-I}
\label{sub:gup1}
Having summarized the standard results for neutrino oscillation probabilities, coherence length and NSI, we proceed to detail the how considered GUPs scenario modify these quantities starting with the GUP-I model in Eq.~\eqref{eq:GUP}.

\subsubsection*{Neutrino oscillations}
Assuming that $m_i \ll p$, $i=1,2,3$, we can expand the dispersion relation in Eq.~\eqref{eq:disp_GUP} as
\begin{align}\label{dispersion-1}
E & \simeq p\sqrt{1 - 2\beta p^2} + \frac{m^2}{2p\sqrt{1-2\beta p^2}}+\mathcal{O}(m^4)\,,
\end{align}
recovering the usual expression in the limit $\beta \rightarrow 0$. Denoting the oscillation frequency in the standard case ($\beta = 0$) as $\Delta^{(0)}$, the GUP-I scenario gives
\be 
\Delta^{(I)}_{ij} = \frac{\Delta^{(0)}_{ij}}{\sqrt{1-2\beta E^2}}\,,
\ee
and the oscillation frequency is consequently increased as the square root is necessarily smaller than unity.

\begin{figure*}
        \centering
        \includegraphics[width=0.82\linewidth]{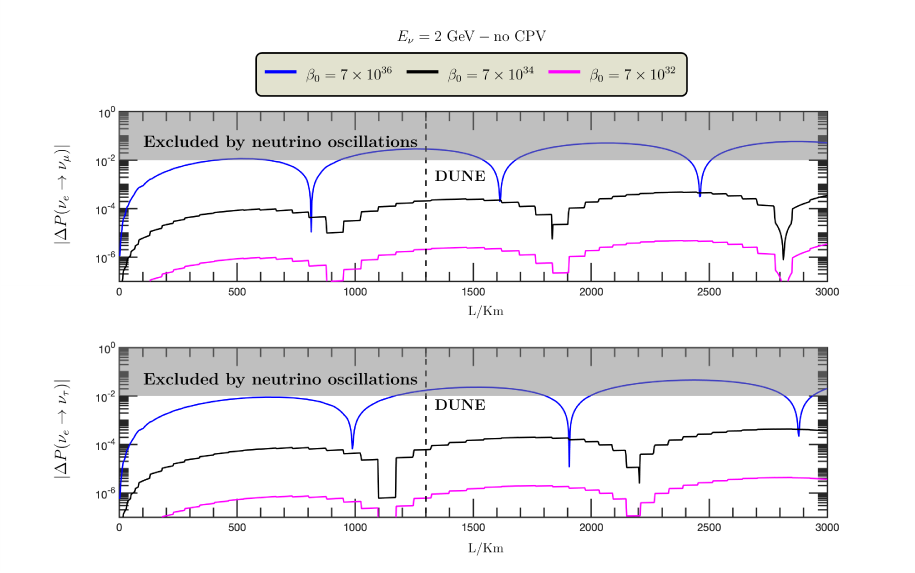}
        \caption{Difference between the transition probabilities obtained for the standard case and in presence of the GUP-I scenario, as a function of the neutrino baseline $L$ and for $E_\nu = 2$ GeV. The dashed vertical lines indicate the DUNE experiment baseline~\cite{DUNE:2020jqi} whereas the gray areas are the exclusion contours determined by the present experimental sensitivity. Top panel: $|\Delta P(\nu_e \rightarrow \nu_\mu)|$. Bottom panel: $|\Delta P(\nu_e \rightarrow \nu_\tau)|$. }
        \label{fig:deltap_1}
\end{figure*}

Consider a case where one oscillation channel is dominant. The related transition probability is given by $(\ell \neq \ell')$
\be 
P_{\ell\ell'} = \sin^2(2\theta)\sin^2\bfrac{\Delta^{(0)}}{\sqrt{1-2\beta E^2}}\,,
\ee 
and assuming $2\beta E^2 \ll 1$ it gives
\begin{align}
P_{\ell\ell'} &\simeq \sin^2(2\theta)(\sin^2 \Delta^{(0)} + 2\Delta^{(0)}\beta E^2\cos \Delta^{(0)})\,.
\end{align}
Therefore, the difference in oscillation probability induced by the GUP is 
\be 
\label{eq:deltapnocoh}
\Delta P = 2\Delta^{(0)} \beta E^2\sin^2(2\theta)\cos \Delta^{(0)} \sim \beta E^2\,.
\ee
The effect is illustrated in Fig.~\ref{fig:deltap_1}, where we plot the change in the $P_{e\mu}$  and $P_{e\tau}$ transitions as a function of the experiment baseline for different values of the GUP parameter and considering a typical neutrino beam energy of 2 GeV.   

\begin{figure}
    \centering
    \includegraphics[width=.95\linewidth]{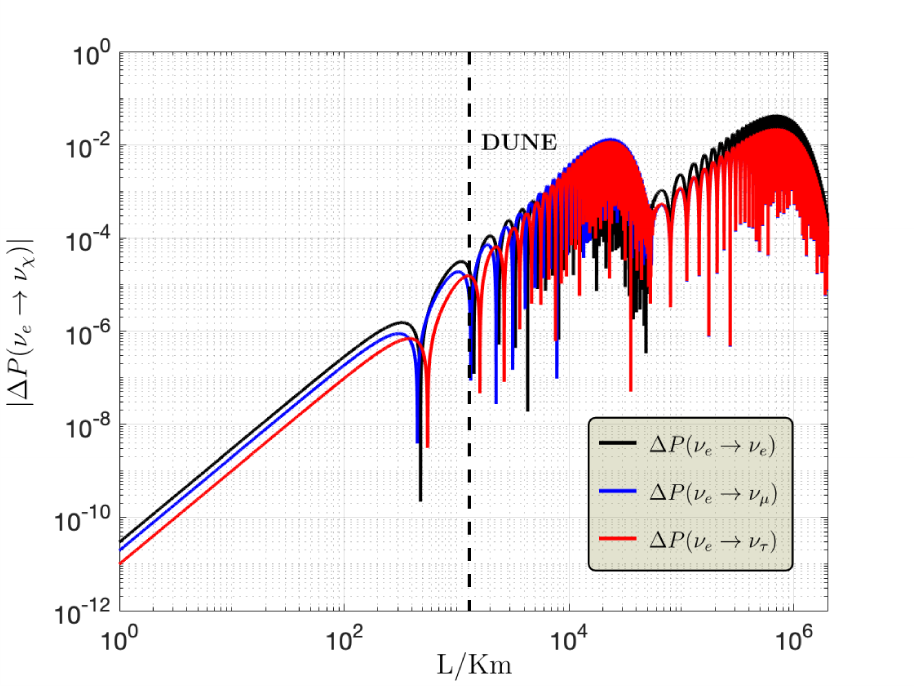}
    \caption{\label{fig:coh}The difference in neutrino oscillation transition probabilities computed for $\beta = 0.01$ GeV$^{-2}$ ($\beta_0 \sim 10^{36}$) and the standard case. For the computation we have assumed normal mass hierarchy, upper $\theta_{23}$ octant, no CPV and set the neutrino energy to $E_\nu = 2$ GeV, which is the expected beam energy of the DUNE experiment with baseline $L=1300 $ km~\cite{DUNE:2020jqi}. We have set the neutrino energy spread to $\sigma_E/E = 0.02.$}
    \label{fig:deltap}
\end{figure}

The current error resolution on transition probabilities, $\Delta P \sim 10^{-2}$, can be used to estimate the maximum GUP effect allowed by current experiments, thereby yielding the exclusion limit indicated by the gray area. The corresponding bound is 
\be 
\beta E^2 = \frac{\beta}{\text{GeV}^{-2}} \times \bfrac{E}{\text{GeV}}^2 \lesssim 10^{-2}\,,
\ee  
and implies the following constraint on the dimensionless  GUP-I parameter:
\be 
\beta_0 \lesssim 10^{36} \times \bfrac{\text{GeV}}{E}^2\,.
\ee
A similar result holds true for the experimental constraints cast by measurements of disappearance oscillation probabilities, as shown from the plot in Fig.~\ref{fig:deltap} for a value $\beta = 0.01$ GeV${}^{-2}$ of the GUP parameter.

\subsubsection*{Nonstandard interactions}
Retaining the first nontrivial order of approximation in the dispersion relation \eqref{dispersion-1},
\be 
E \simeq p\sqrt{1 - 2\beta p^2} + \frac{m^2}{2p}(1-2\beta p^2)^{-1/2},
\ee 
we can write the Hamiltonian for neutrino oscillations:
\be 
H = \frac{1}{2E}U^\dagger\begin{pmatrix}
    0 & 0 & 0 \\ 0 & \Delta m_{21}^2 & 0 \\ 0 & 0 & \Delta m_{31}^2
\end{pmatrix}U(1+\beta E^2)\,,
\ee 
where the perturbation $\beta E^2 \ll 1$ is the NSI term due to GUP. By solving via matching for $\beta$ and taking the PMNS matrix elements to be $\mathcal{O}(1)$, we find 
\be 
\beta = \frac{2\sqrt{2}G_FN_e\varepsilon}{E\Delta m_{ij}^2}\,,
\ee 
approximately yielding for the magnitude of $\beta$
\be
|\beta|  \sim \frac{N_e}{N_A/\text{cm}^3}\times \frac{10^{-5} \text{ eV}^2}{\Delta m_{ij}^2} \times \frac{|\varepsilon|}{10^{-2}} \frac{\text{GeV}}{E}\times 10^{-19} \text{ eV}^{-2}\,,
\ee 
or, in terms of the dimensionless quantity $\beta_0$,
\be
 \beta_0 \lesssim 10^{37}\times  \frac{\text{GeV}}{E}\,.
\ee
Although experiments utilizing lower energy neutrinos seem more suitable for constraining $\beta_0$ through the related NSI, the present experimental sensitivities are far from casting stringent bounds. For instance, the COHERENT~\cite{COHERENT:2017ipa,COHERENT:2020iec,Giunti:2019xpr,Karkkainen:2023ozw} and BOREXINO~\cite{Borexino:2019mhy} experiments yield $|\varepsilon| \lesssim 10$ and essentially result in no bound on the GUP parameter of interest. However, the future experiments Hyper-K \cite{Hyper-Kamiokande:2018ofw} and JUNO \cite{JUNO:2021vlw} are expected to probe neutrino oscillation parameters related to the solar flux---with neutrino energies of a few MeV--- reaching a combined sensitivity to NSI that would constrain $|\varepsilon| \lesssim 0.1$, corresponding to $\beta_0 \lesssim 10^{35}$.

\subsubsection*{Coherence length}
GUPs also modify the neutrino coherence length as it is directly sensitive to the uncertainty in the neutrino position. For the GUP-I scenario, considering terms beyond the plane wave approximation, we have 
\be\label{eq:cohrel}
L^\text{coh,(I)}_{ij} = \frac{L_{ij}^\text{coh,(0)}}{1 + 3\beta E_\nu^2}\,,
\ee
where $\Delta x$ is the neutrino position uncertainty and $L_{ij}^\text{coh,(0)}$ the standard coherence length in Eq.~\eqref{eq:cohele}. The latter is clearly recovered in the limit $\beta \rightarrow 0$.

The IceCube neutrino observatory~\cite{ICECUBE:2024fej} has recently given an upper bound on the decoherence parameter $\Gamma$, which translates into a lower bound on the decoherence length $L^\text{coh} = c\hbar/\Gamma$ and so directly probes the quantity in Eq.~\eqref{eq:cohrel}. In particular, utilizing $E_\nu=E_\nu^\text{peak} = 1$ TeV and $\Delta E/E = 0.69$, the 90\% confidence level IceCube bound gives
\be 
\beta_0 \lesssim 10^{30}.
\ee 
Since IceCube is designed to detect astrophysical high-energy neutrinos, the neutrino energy distribution peak is much higher than in more traditional neutrino oscillation experiments. Given a lower bound on the coherence length, from Eq.~\eqref{eq:cohrel} it is evident that this increase in energy helps to cast a stricter bound on the GUP parameter $\beta$.

\begin{figure}
    \centering
    \includegraphics[width=.95\linewidth]{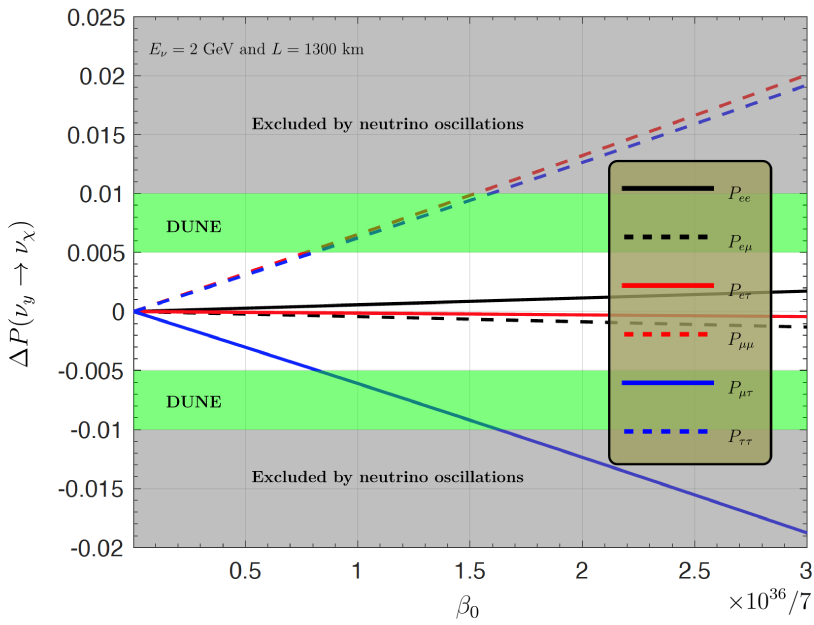}
    \caption{Differences between the standard transition probabilities and the corresponding quantities obtained in the GUP-I scenario as a function of the dimensionless  parameter $\beta_0$. The plot uses the DUNE experiment baseline and neutrino energy. The gray areas are the exclusion contours determined by the present experimental sensitivity, the green bands indicate instead the bounds obtainable with the expected DUNE sensitivity. }
    \label{fig:dune}
\end{figure}

As for the difference in the transition probability induced by GUP through the change in the coherence length,
we find
\be 
\Delta P \sim \beta E^2 \bfrac{L}{L^{\text{coh},(0)}}^2\,,
\ee
in the limit $\beta E^2 \ll 1$ and $L \ll L^{\text{coh}^{(0)}}$. Although the effect is subdominant with respect to the contribution in Eq.~\eqref{eq:deltapnocoh}, we include it in the computation of the transition probability differences shown in Fig.~\ref{fig:dune}. In the plot, we set baseline and neutrino beam energy to the values proper of the DUNE experiment and analyze the power of the latter to constrain the dimensionless  parameter $\beta_0$. As we can see, the bound obtained (gray exclusion area) is in line with the results obtained for neutrino oscillations in the plane wave limit, confirming that changes in the coherence length yield subdominant effects at this baselines. The green bands indicate the exclusion bounds obtainable with the foreseen sensitivity of the DUNE experiment~\cite{DUNE:2020jqi}.

\subsection{GUP-II}
\label{sub:gup2}
We now repeat the previous analysis for the second GUP scenario considered, specified in Eq.~\eqref{eq:Heis_alg_2}.  

\subsubsection*{Neutrino oscillations}
Expanding, in this case, the dispersion relation~\eqref{eq:disp_GUP_2} in the limit $\alpha p \ll 1$ we obtain
\be 
E = p|1-\alpha p| + \frac{m^2}{2p}(1 + \alpha p)+\mathcal{O}(m^4)\,,
\ee 
corresponding to an oscillation frequency for the GUP-II case of
\be 
\Delta_{ij}^{(II)} = \frac{\Delta_{ij}^{(0)}}{|1-\alpha E|}\,.
\ee 
Similarly to GUP-I case, in two flavour approximation the transition probability for $\nu_\ell \rightarrow \nu_{\ell'}$ oscillation is given by
\be 
P_{\ell\ell'} = \sin^2(2\theta)\sin^2\left(\frac{\Delta^{(0)}}{|1-\alpha E|}\right)\,.
\ee 
With the further assumption $\alpha E  \ll 1$, the difference in oscillation probabilities between the standard and the GUP-II relations can be written as
\be 
\label{eq:DP22}
\Delta P = 2\Delta^{(0)}\alpha \sin^2(2\theta)E\cos \Delta^{(0)} \sim \alpha E\,.
\ee 
The current experimental resolution proper of transition probability measurements, $\Delta P \sim 10^{-2}$, implies
\be 
\alpha E = \frac{\alpha}{\text{GeV}^{-1}}\times \frac{E}{\text{GeV}} \lesssim 10^{-2}\,.
\ee 
Extracting the typical scale that appears in the definition of $\alpha$, $\a  \simeq 8 \a_0 \times10^{-20} \GeV^{-1}$ the above equation can be recast as a constraint for the dimensionless variable $\alpha_0$:
\be 
\alpha_0  \lesssim 10^{17}  \times \frac{\text{GeV}}{E}\,.
\ee 

\subsubsection*{Nonstandard interaction}
Retaining again only the first terms in the dispersion relation expansion
\be 
E = p|1-\alpha p| + \frac{m^2}{2p}(1 + \alpha p)\,,
\ee 
we proceed as in the GUP-I case using the same-momentum approximation and disregarding terms proportional to the unit matrix. The relevant Hamiltonian describing neutrino oscillations is then given by
\be 
H = \frac{1}{2E}U^\dagger\begin{pmatrix}
    0 & 0 & 0 \\ 0 & \Delta m_{21}^2 & 0 \\ 0 & 0 & \Delta m_{31}^2
\end{pmatrix}U(1 + \alpha E)\,,
\ee 
in which the extra term proportional to $\alpha$ can be interpreted as a matter NSI term. Via matching, we find
\be 
\alpha = \frac{2\sqrt{2}G_FN_e\varepsilon}{\Delta m_{ij}^2}\,,
\ee 
\FloatBarrier
\noindent where we take the PMNS matrix elements to be $\mathcal{O}(1)$. We notice that \textit{in the context of NSI}
\be 
\alpha = \beta E\,,
\ee 
and since $\beta \sim E^{-1}$, the expression for the GUP-II parameter $\alpha$ does not involve the neutrino energy. We estimate
\be
|\alpha|  \sim \frac{N_e}{N_A/\text{cm}^3}\times \frac{10^{-5} \text{ eV}^2}{\Delta m_{ij}^2} \times \frac{|\varepsilon|}{10^{-2}} \times 10^{-10} \text{ eV}^{-1}\,,
\ee 
and obtain the following order-of-magnitude approximation for the current experimental constraint on the GUP-II parameter $|\alpha| \lesssim10^{-10}$ eV$^{-1}$ or, equivalently:
    \be 
     %\alpha\bfrac{\text{GeV}}{M_P}^2 \times \bfrac{E}{\text{GeV}}^2 \leq \mathcal{O}(10^{-2}) \Rightarrow
     \a_0 \lesssim 10^{18}\,.
    \ee 
   
\subsubsection*{Coherence length}
The coherence length resulting from the GUP-II scenario is
\be 
L_{ij}^\text{coh,(II)} = \frac{L_{ij}^\text{coh,(0)}}{1 + 2\alpha E_\nu}\,,
\ee 
and correctly recovers the standard case for  $\alpha \rightarrow 0$.

Similar to the previous analysis, we find that the IceCube data results in a strong bound on $\alpha_0$
\be 
\alpha_0 \lesssim 10^{16}\,.
\ee 
The difference in transition probability induced by the change in the coherence length is instead given by
\be 
\Delta P \sim \alpha E \bfrac{L}{L^{\text{coh},(0)}}^2\,,
\ee
in the limit $\alpha E \ll 1$ and $L \ll L^{\text{coh},(0)}$. As we can see, also for GUP-II scenario the effect is subdominant with respect to that in Eq.~\eqref{eq:DP22} at current experiments.

%%%%%%%%%%%%%%%%%%%%%%%%%%%
%% IMPORTANT NUMBERSSSSSS
% Beta0 = 0.8*1e-3, DUNE
% Beta0 = 0.21, HK (262.5 times DUNE)
% Beta0 = 71, JUNO (88750 times DUNE)

% Alpha = 1.6*1e-3, DUNE
% Alpha = 0.065, HK
% Alpha = 0.17, JUNO
% Alpha0 = 1e20/8*alpha
%
%%%

\section{Results}
In this letter we have investigated the experimental feasibility of using the low energy neutrino data to probe the phenomenological effects of two GUP scenarios. Whereas experiments are not sensible enough to distinguish between the two cases analyzed, we find that the data can indeed constrain the involved GUP parameters. In Tab.~\ref{tab:1} we collect the constraints obtained from the results in Secs.~\ref{sub:gup1} and~\ref{sub:gup2} considering the specifics of three upcoming neutrino experiments, Hyper-Kamiokande~\cite{Hyper-Kamiokande:2018ofw}, DUNE~\cite{DUNE:2020jqi} and JUNO~\cite{JUNO:2021vlw}, and the recent constraint on the decoherence parameter by the IceCube neutrino observatory~\cite{ICECUBE:2024fej}. For oscillation experiments, the quoted bounds are mainly driven by the modifications that the considered GUP scenario yield on the neutrino oscillation pattern computed in the plane wave approximation. These constraints are comparable to the---marginally milder---exclusions obtained by modeling the GUP effects as nonstandard neutrino interactions and strongly outclass the limit obtained from the changes in the oscillation pattern induced by the modified neutrino coherence length. The latter is directly probed by the IceCube result, which yields the tightest constraint on the GUP parameters owning also to the higher peak energy of the observed neutrino spectrum. The energy dependence of the mentioned bounds reflects the fact that the modifications to the uncertainty principle in the considered GUP scenarios are necessarily more prominent at higher energies or, equivalently, at  larger momenta for set particle masses. Consequently, the upper bounds on GUP parameters scale inversely with the energy even if the parameters themselves are constants.  

\begin{table}[h]
        \begin{tabular}{c|cccc|ccc}
            \toprule
             Experiment & $L$/km & $E_\nu^\text{peak}$/MeV & $\Delta P$ && $\alpha_0$ & &$\beta_0$ \\
            \midrule
             DUNE~\cite{DUNE:2020jqi} & 1300 & 2000 & $ 0.5$ \% & &$ 10^{16}$ && $10^{36}$\\
             Hyper-K~\cite{Hyper-Kamiokande:2018ofw} & 295 & 600 & $ 1$ \% &&$ 10^{18}$ &&$ 10^{38}$\\
             JUNO~\cite{JUNO:2021vlw} & 53 & 4.5 & $ 1$ \% & &$ 10^{18}$ &&$ 10^{41}$\\
             ICECUBE~\cite{ICECUBE:2024fej} & $-$ & $10^6$ & $-$ && $10^{16}$ && $10^{30}$\\
             \bottomrule
        \end{tabular}
        \caption{First three lines: the future neutrino oscillation experiments, their baselines, the most probable neutrino energies and foreseen measurement accuracy. The last two columns shown the achievable upper bounds on the GUP-parameters $\alpha_0$ and $\beta_0$. The last line shows the exclusion obtained with high energy neutrinos at the IceCube neutrino observatory~\cite{ICECUBE:2024fej}.  }
        \label{tab:1}
\end{table}

To put our results into perspective, we refer the reader to the recent review paper~\cite{Bosso:2023aht} and the tables contained, which list the upper bounds on the parameter $\beta_0$ derived from various gravitational and non-gravitational experiments and observations. As we can see, the considered neutrino experiments~\cite{Hyper-Kamiokande:2018ofw,DUNE:2020jqi,JUNO:2021vlw} and IceCube~\cite{ICECUBE:2024fej}, in particular, impose the strictest bounds on the GUP parameters.

\acknowledgments
This work was supported by the Estonian Research Council grants  MOB3JD1202, PRG803, RVTT3,  RVTT7, and by the CoE program TK202 ``Fundamental Universe''.

\bibliography{Bib_on_GUP}

\end{document}